\documentclass[manuscript,screen]{acmart}

\AtBeginDocument{%
  \providecommand\BibTeX{{%
    \normalfont B\kern-0.5em{\scshape i\kern-0.25em b}\kern-0.8em\TeX}}}

\setcopyright{acmcopyright}
\copyrightyear{2018}
\acmYear{2018}
\acmDOI{XXXXXXX.XXXXXXX}

\acmConference[Conference acronym 'XX]{Make sure to enter the correct
  conference title from your rights confirmation emai}{June 03--05,
  2018}{Woodstock, NY}
\acmPrice{15.00}
\acmISBN{978-1-4503-XXXX-X/18/06}

\usepackage{stfloats}
\usepackage{url}
\usepackage{caption}
\usepackage{subcaption}
\usepackage{tikz}
\usepackage{graphicx}
\usepackage{amsmath}
\usepackage{amsfonts}

\usepackage{amssymb}
\usepackage{mathtools}
\usepackage{multirow}
\usepackage[ruled,vlined,linesnumbered]{algorithm2e}

\usepackage[utf8]{inputenc}
\begin{document}

\title{Collision Aware Data Allocation In Multi-tube DNA Storage}

\author{Yixun Wei}
\email{Yixun09@gmail.com}
\affiliation{%
  \institution{Department of Computer Science and Engineering, University of Minnesota}
   \country{USA}
}

\author{Bingzhe Li }
\affiliation{%
 \institution{Department of Electrical and Computer Engineering, Oklahoma State University}
  \country{USA}}

\author{David H.C. Du}
\affiliation{%
  \institution{Department of Computer Science and Engineering, University of Minnesota}
   \country{USA}}

\renewcommand{\shortauthors}{Yixun Wei, et al.}

\begin{abstract}
DNA storage is a promising archival data storage solution to today’s big data problem. A DNA storage system encodes and stores digital data with synthetic DNA sequences and decodes DNA sequences back to digital data via sequencing. For efficient target data retrieving, existing Polymerase Chain Reaction (PCR) based DNA storage systems apply primers as specific identifiers to tag different sets of DNA strands. However, if a primer has collisions with any payload in the same DNA tube, the primer cannot safely serve as an identifier and must be disabled in this tube. In a DNA storage system with multiple DNA tubes, the primer-payload collisions can spread over all DNA tubes, repeatedly disable many primers, and cause a significant overall capacity reduction. 

This paper proposes using a collision-aware data allocation scheme to allocate data with different collisions into different tubes so that a primer banned in a tube because of primer-payload collision can be reused in other tubes. This allocation helps increase the number of usable primers over all tubes thus enhancing the overall storage capacity. The executing time of our scheme is $O(n^2)$ to the number of digital data chunks. The scheme serves as a pre-processing method for any DNA storage system. The evaluation of the state-of-the-art encoding scheme shows that the scheme can increase 20\%-25\% overall storage capacity.   
\end{abstract}



\received{xx February xxxx}
\received[revised]{xx March xxxx}
\received[accepted]{xx June xxxx}

\maketitle

\section{Introduction}
Modern archival storage has been trying to preserve the ever-growing digital data reliably for centuries~\cite{miller2020future}. However, typical archival storage media/devices are not matching the booming storage demand. The storage requirement in hyper-scale data centres is expected to reach 32.6 million petabytes in 2030, which might surpass the total supplied storage capacity~\cite{alliance2021preserving}\cite{IDC}. Besides, the typical storage media/devices are also not durable enough (usually within a decade), thus introducing expensive data preservation costs.

DNA is emerging as a promising archival storage medium to meet the burgeoning storage demand. Theoretically, DNA storage can have a density of about 1 exabyte/mm$^3$  and preserve the data for hundreds of years~\cite{bornholt2016dna}. Even considering the implementation overheads of a practical archival system, existing DNA storage systems can still be orders of magnitude denser than tape\cite{alliance2021preserving}. Additionally, other unique properties of DNA, such as sustainability and energy efficiency, further make DNA a desirable medium for storing archival data for many decades.

In existing DNA storage systems, digital data is encoded and synthesized as the payloads of DNA strands. The DNA strands are usually dehydrated and stored in physical tubes for long-term preservation. When retrieving data, people liquidize the DNA and use a drop of the liquid for sequencing (i.e., the read process in DNA storage). To ensure efficient data retrieval, polymerase chain reaction (PCR) based random access has been introduced to DNA storage~\cite{organick2018random}\cite{bornholt2016dna}\cite{yazdi2015rewritable}\cite{yazdi2017portable}. In the PCR-based random access, each DNA payload is flanked by a pair of primers (i.e., short nucleic acid sequences) to form DNA strands. The primers serve as unique tags to identify different sets of DNA strands. When retrieving payloads (data), only strands with specific primer pairs can be amplified by PCR and then sequenced. In such a system, the number of DNA payloads stored in a tube is proportional to the number of usable primers in a tube.

The existence of primer-payload collisions disables many primers. Primer-payload collisions refer to almost identical subsequences between a primer and any portion of a payload~\cite{organick2018random}. A primer must be disabled in the tube if it collides with any payloads in the same tube. Otherwise, the PCR may amplify wrong DNA sequences and cause sequencing failure. Wei et.al.~\cite{wei2022dna} investigated that the number of usable primers in a single tube can decrease up to 70\%-99\% as the number of payloads increases. As a result, the single tube capacity of random-access-based DNA storage is significantly decreased.


Because a primer can be reused among different tubes (as long as this primer has no collision in the tubes), the overall storage capacity in a multi-tube DNA storage system is also greatly affected by the primer-payload collisions. In this paper, we first expand the investigation of DNA storage capacity from single tube to multiple tubes. We find the overall storage capacity of multiple tubes decreases greatly as more than 70\% \& 99\%  primers in the primer library are disabled in all tubes for encoding schemes rotation~\cite{bornholt2016dna} and blawat~\cite{blawat2016forward} respectively. This is because the data is sequentially allocated to one tube after another. The same primer-payload collisions spread over almost all tubes and each tube has to suffer from the capacity reduction.

This paper proposes a collision aware data allocation scheme to allocate digital data chunks with different collisions into different tubes so that a primer disabled in one tube can be still reusable in other tubes. The collision aware data allocation consists of an initial clustering and a following refinement. The initial clustering tries to find size-restricted clusters so that data chunks inside each cluster have a small union of colluded primers. Different from a typical clustering which directly calculates objects' distance and groups objects based on their distance, the initial clustering requires comparing the collided primer union among data chunks. We adopt a hierarchical clustering with a specific clustering metric "merge priority" to monitor the collided primer union among data chunks and existing clusters. We then progressively merge small clusters to big clusters until all clusters are too large to be further merged. The clusters correspond to DNA tubes. We further fine-tube data chunks in each cluster to fully use up all space. Finally data chunks in different clusters will be assigned to different tubes. Note that the collision aware data allocation scheme, which serves as a pre-processing method, is independent from other parts of DNA storage system (e.g., encoding). We can apply it to any existing DNA storage system to gain storage capacity enhancements.

The rest of this paper is organized as follows. Section~\ref{sec:multitube backgroud} introduces DNA storage backgrounds, including storage workflow, factors that affect capacity, primer-payload collision, and overall multi-tube DNA storage capacity. Section~\ref{sec:multitube solution} states the data allocation problem and elaborates on the collision aware data allocation scheme. Section~\ref{sec:multitube evaluation} evaluates our scheme. Conclusions are drawn in Section~\ref{sec:multitube conclusion}.

\section{Background}

\label{sec:multitube backgroud}
\subsection{DNA Storage in Brief}

\begin{figure*}[htbp]
\centering
\small
\includegraphics[width=\textwidth]{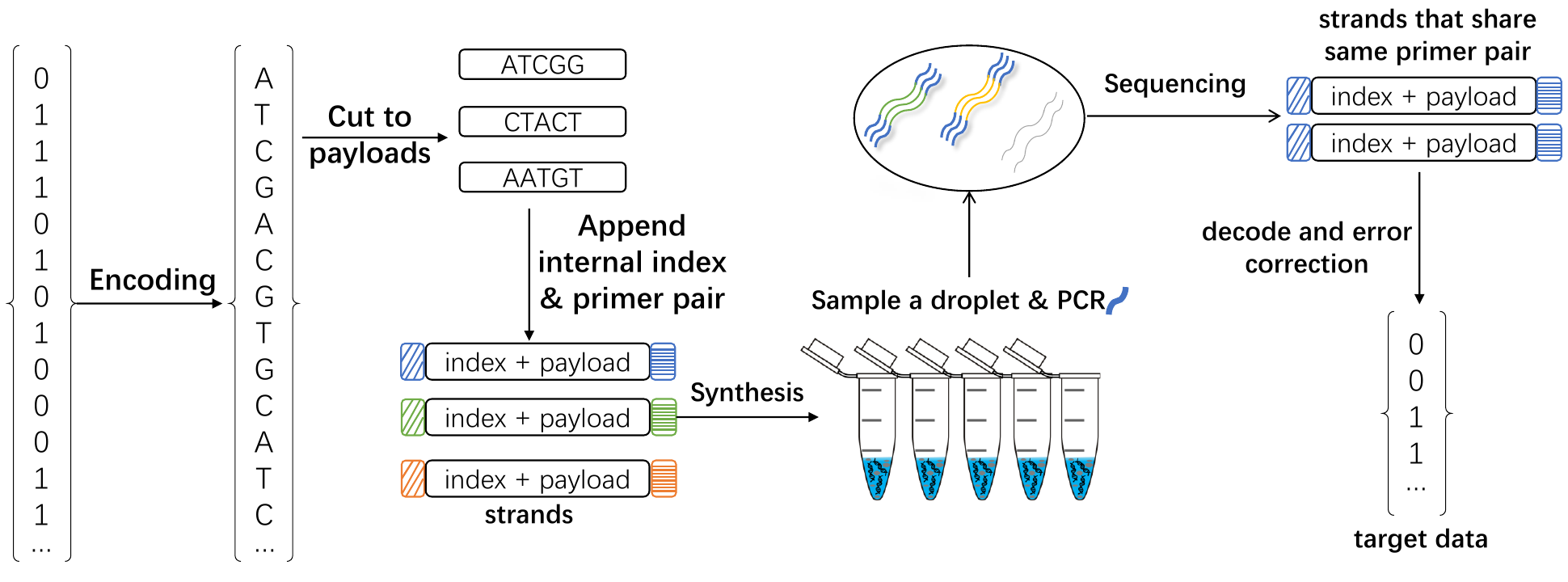}
\caption{Workflow of typical DNA storage system}
\label{fig:DNA storage workflow}
\end{figure*}

Several researchers have demonstrated the feasibility and discussed the potential capacity of DNA archival storage~\cite{DNAfuture}~\cite{li2020can}\cite{wei2022dna}. Figure~\ref{fig:DNA storage workflow} shows the typical workflow of the current DNA storage system.

In existing DNA storage systems, digital data will first be added with some error correction code (e.g., Reed–Solomon Code) for data recoverability~\cite{lin2022managing}~\cite{carmean2018dna}. Certain encoding schemes will then encode the resultant data to a sequence of DNA bases (A, T, G, and C). The DNA sequences will be cut into multiple shorter subsequences called DNA payloads. Each payload will be flanked by a pair of primers and an internal index to form a DNA strand. The primer pair is a unique tag that enables random access to DNA storage based on PCR. Multiple DNA strands can share one primer pair, and an internal index helps identify the strands. Eventually, each DNA strand will be chemically synthesized base by base and stored in physical tubes.

\begin{figure}[htbp]
  \begin{subfigure}[b]{0.45\textwidth}
    \includegraphics[width=\textwidth]{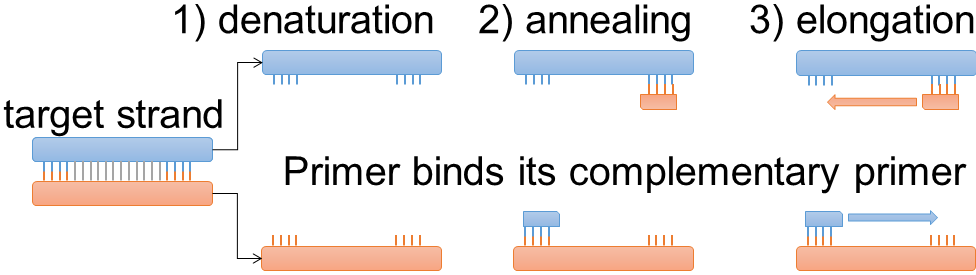}
    \caption{standard PCR}
    \label{fig:standard PCR}
  \end{subfigure}
  \hfill
  \begin{subfigure}[b]{0.45\textwidth}
    \includegraphics[width=\textwidth]{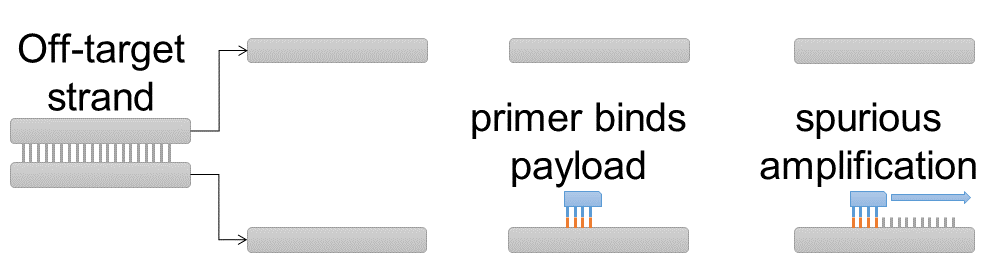}
    \caption{PCR with primer-payload collision} 
    \label{fig:primer-payload collision}
  \end{subfigure}
  \caption{standard PCR and defective PCR with primer-payload collision}
\end{figure}

The PCR-based random access is a process that first amplifies target DNA strands with a specific primer pair and then sequences only the target strands. The amplification usually takes 20 to 40 PCR cycles. A standard PCR cycle is shown in Figure 2 (a). In each cycle, target DNA strands are denatured into two single-stranded DNA molecules, and each molecule anneals with the added complementary primer templates. Then the DNA bases (A, T, C, G) from the added mixture solution will pair with the annealed molecules to form new DNA strands. After multiple PCR cycles, the target strands have enough concentration and can be sequenced out. Finally, the read-out DNA sequences can be decoded back to digital data. The ECC added before can help recover some errors during the synthesis and sequencing processes.

\subsection{Primer-Payload Collisions}
\label{sec:tube capacity}
A DNA storage system must obey many bio-constraints to provide reliable data storage; thus, many factors affect DNA tube capacity. DNA tube capacity is the multiplication of the following factors~\cite{li2020can}\cite{wei2022dna}: 1) DNA payload length, 2) the number of digital bits each DNA base can store (i.e., encoding density), 3) the number of strands that one primer pair can accommodate (i.e., parallel factor), and 4) the number of usable primers. A practical DNA strand length is limited to 100-300 bases because the error rate soars up when the strand becomes overlong~\cite{alliance2021preserving}\cite{matange2021dna}. After excluding the primer pair and internal index, a payload is usually no longer than 200 bases. Similarly, the parallel factor is also limited (e.g., 1.55 $\times 10^6$~\cite{li2020can}). Based on different encoding schemes, the encoding density varies from less than one bit per DNA base to at most two bits per DNA base (e.g., A=00, T=01, C=10, G=11). The number of usable primers is also limited by certain primer design rules and the primer-payload collisions that we will discuss later. The DNA tube capacity equals the payload length $\times$ encoding density $\times$ parallel factor $\times$ the number of usable primers / 2 (i.e., the number of primer pairs).

Given all biological constraints, primer-payload collision is the most critical factor affecting the tube capacity. As Figure 2 (b) shows, a collision occurs if a pair of almost identical subsequences exists between a primer and a portion of payload stored in the same tube~\cite{organick2018random}\cite{wei2022dna}. The subsequences are usually longer than 12 bases and have at most two mismatches or gaps. In each PCR cycle, the PCR can amplify some irrelevant strands if these strands contain payloads that collide with the target primer. More importantly, this consumes the limited PCR reagents (e.g., complementary primers used to bind with the target primer and free DNA bases waiting for complementing new double-helix strands). In this case, PCR amplification is very vulnerable to primer-payload collisions. The target primer must compete for the limited PCR reagents with thousands of collided payloads in each PCR cycle. Even if a primer has only a few collisions at the beginning, the collisions can still significantly impact the final result. That is because PCR is an exponential augmentation process. Slight amplification variations can exacerbate cycle by cycle and consume ever more PCR reagents. Eventually, the resultant solution will be saturated with plenty of irrelevant sequences but very few target sequences. As a result, the sequencing for target sequences is inhibited.

Organick et al. first ~\cite{organick2018random} reported a PCR failure due to primer-payload collisions. They successfully accessed a target file when the DNA tube contained no other data but failed to do so when there were nine files in the tube (the target data is 17.4\% of the whole data in the tube). After that, Wei et al.~\cite{wei2022dna} investigated the impact of primer payload collisions based on 1.5TB of different types of digital data. Their results showed that the primer payload collision could cause 70\% to 99\% tube capacity loss depending on different encoding schemes. Most of the existing encoding schemes can achieve only several Gigabytes of tube capacity, which is much lower than people's expectations and may inhibit DNA storage from practical use.

\section{Collision Aware Data Allocation}
\label{sec:multitube solution}
The last section has shown that sequentially assigning data tube by tube will cause the primer-payload collisions to spread over all tubes, causing significant capacity reduction on all tubes. In this section, we propose a collision aware data allocation to restrict the distribute of primer-payload collisions. We first state the problem definition of collision aware data allocation. We then describe the overall procedure of the data allocation scheme. After that, we explain the allocation scheme step by step.

\subsection{Problem Statement}
\label{sec:multitube problem statement}
The main idea of the collision aware data allocation is assigning data chunks into different tubes based on their collided primers so that one collided primer is only disabled in a few number of tubes rather than all tubes. This main idea sounds similar to a typical clustering but is different from the clustering criteria. In typical clustering, the "similarity" among objects can be directly calculated (e.g., the Cosine distance of two vectors). While in the collision aware data allocation, the "similarity" among objects is the increment of the union of two data chunks' collided primers. Two data chunks with totally different sets of collided primers can be in the same cluster if the cluster has already included both the two data chunks' collided primers. That is because adding the two data chunks to the cluster does not introduce extra collided primers.

We first define two properties of data chunks and clusters.
\begin{itemize}	
    \item \textit{Chunk/Cluster collided primers:} A data chunk's collided primers are the collection of primers that have collisions with the data chunk. A cluster's collided primers are the collection of primers that have collisions with data chunks inside the cluster.	
    	
    \item \textit{Cluster/Chunk size:} Archival storage usually stores digital data as a sequence of data chunks. The size of chunks varies among different systems, we assume the data chunk size is 4 KB in this paper. The size of a cluster is the summation of the data chunk size in the cluster. In the collision aware data allocation, a cluster is corresponding to a DNA tube. The size of the cluster should not exceed the tube capacity which is determined by the number of remaining usable primers in the cluster.
\end{itemize}

\noindent\textbf{Goal:} Given all data chunks $C=(c_1,c_2,....c_{|C|})$ that belong to different files/objects, partition $C$ into multiple groups $G=(g_1,g_2...)$ so that $\sum_{g_i \in G} |g_i\textit{'s collided primers}|$ is minimum. 

\BlankLine

\noindent In reality, all data chunks in a group will be assigned to a DNA tube. The size of chunks cannot exceed the capacity of the DNA tube which is determined by the number of usable primers. Also, because data chunks belong to different files/objects, we do not want one file's data chunks scattered over too many different places and introduce too much sequencing to retrieve a file. 

\noindent\textbf{Constraints:} 
For each $g_i$ the $\sum_{c_i \in g_i} \textit{chunk size}(c_i) < \sum_{\textit{primers } \notin \textit{ } g_i \textit{'s collided primers}} \textit{primer capacity}$. When retrieving all chunks of a file/object, the number of sequencing should be restricted to a certain number (e.g., K=5).

\subsection{Overall Procedure}
Assuming we have the archival data in the form of data chunks (4KB), we first encode them to DNA strands and check the collision of each data chunk.
With the knowledge of what primers each data chunk has collisions with, we propose a collision aware data allocation which consists of two steps:  1) initial clustering, and 2) refinement.
The initial clustering automatically partitions all data chunks into multiple groups (the number of groups is determined by the clustering).
Each group corresponds to a DNS tube.
The collided primers of data chunks inside a group are as similar as possible, while the collided primers of data chunks among different group is dissimilar.
Because the number of collided primers in each group is different, the tube capacity of each group's corresponding tube is also different.

After the initial clustering, we choose the group that is closest to its tube capacity and keep migrating data chunks from other groups to the selected group until its size reaches its tube capacity.
The maximum size of a group is the maximum size of a tube which is determined by the number of usable primers.
We migrate the data chunk whose collided primers are closest to the collided primers in the group so that the migration can induce the least number of new collided primers.
Each time we select one group and fill it up.
All chunks in the group will be assigned to a DNA tube to store.
We then redo the clustering to partition all remaining chunks again and select the next group to fill until all data chunks are assigned to a tube.
Compared with sequentially assigning data chunks into tubes, our allocation can improve 20\% to 25\% of average tube capacity. 

The above allocation procedure ensures a capacity increase but does not pay attention to the potential sequencing overhead. As data chunks are allocated to different places, retrieving one file may need to sequence multiple primer pairs causing an increase in the sequencing overhead. To restrict the sequencing overhead, for all data chunks assigned in one tube, we assign them to different primer pairs based on which file they belong to. We assign chunks from the same file to the same primer pairs so that all data chunks of a file in a tube will be sequenced out in one shot. Consequently, retrieving a file needs sequencing with at most the number of total DNA tubes.  

\subsection{Initial Clustering}
\label{sec:initial clustering}
In the beginning, each data chunk itself is a cluster.
Every cluster has its size and collided primers.
The cluster size is the total size of data chunks inside the cluster and the cluster collided primers are the union of internal chunks' collided primers.
A cluster corresponds to a DNA tube.
The maximum cluster size is the corresponding DNA tube capacity which is determined by the number of collided primers.
Therefore, our goal is to decrease the overall number of collided primers among all clusters.
Chunks in different clusters should have distinct sets of collided primers.
The clustering method keeps merging small clusters similar in terms of the collided primers until all clusters are too large to be further merged.
A cluster is too large to merge if any merging will make the cluster size exceed its corresponding tube capacity.

\begin{figure}[htbp]
    \centering
        \includegraphics[width=1.1\textwidth,height=9cm]{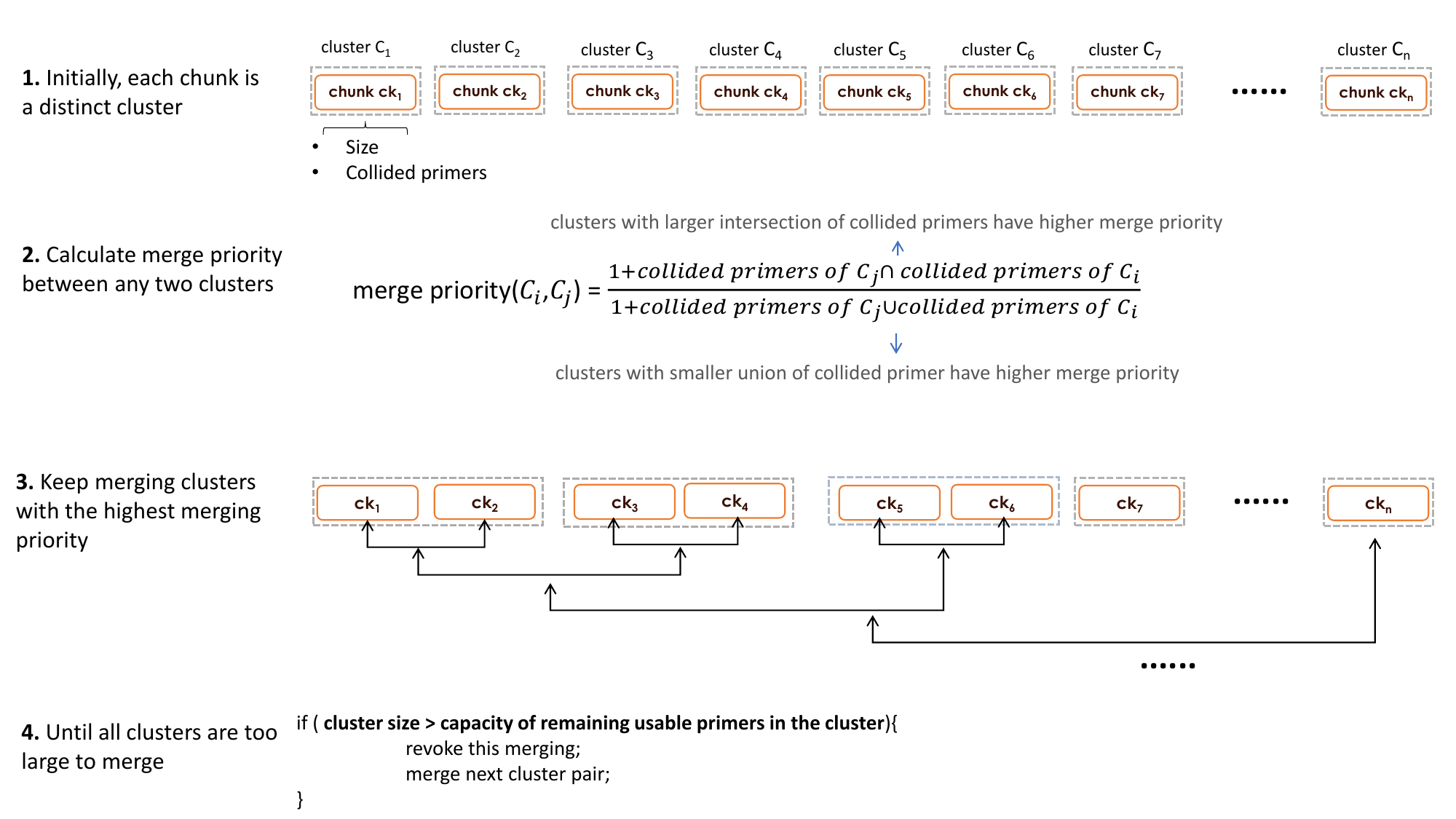}
        \caption{Procedure of Initial Clustering}
        \label{fig:clustering}
\end{figure}

We define a \textit{merge priority} between every two clusters as $\frac{1+\textit{collided primers of }c_i \cap \textit{collided primers of }c_j}{1+\textit{collided primers of }c_i \cup \textit{collided primers of }c_j}$. The procedure of initial clustering is shown in Figure~\ref{fig:clustering}. We calculate the merge priority between every two clusters and keep merging clusters with the highest merging priority until all clusters are too large to merge. Each time we merge a new cluster, we only need to recalculate the merge priority between the new cluster and others.

As discussed in section~\ref{sec:multitube problem statement}, the clustering should not simply partition data chunks based on the similarity of collided primers. Since the goal is to find clusters that each cluster has a small union of collided primers, the clustering should monitor the collided primer unions among different clusters. We select hierarchical clustering as the initial clustering in which the clustering procedure is visually represented as a hierarchical tree: data chunks and small clusters are merged into a bigger cluster at a certain level of the hierarchical tree. This hierarchical and progressive merging process is helpful in monitoring the collide primer unions between existing clusters. Every merging will add the chunk (or small cluster) to a cluster that has the minimum union of collided primers.

There are several typical similarity metrics used in hierarchical clustering. For example, UPGMA~\cite{garcia2009dendroupgma} calculates the distance of all data chunk pairs in two clusters and uses the average distance as the distance of the two clusters (i.e., $D(C_i,C_j)=\frac{1}{\vert \vert C_i \vert \vert \vert \vert C_j \vert \vert}\sum_{x \in C_i,y \in C_j}d(x,y)$). However, these typical similarity metrics are still based on the similarity between data chunks. We use \textit{merge priority} to replace the typical metrics so that the clustering process can better measure the union of merging two clusters. In Section~\ref{sec:multitube evaluation} we will compare the difference between the merge priority and UPGMA.

\subsection{Refinement}
After the initial clustering, we select a cluster that is closest to its tube capacity. If the cluster size is still smaller than its corresponding tube capacity (i.e., size of chunks in cluster $<$ capacity of remaining usable primers), we migrate data chunks from other clusters to the selected cluster. The selected chunks should have the highest merge priority with the selected cluster. After this cluster is fully filled. All the chunks inside are assigned to a DNA tube. For those chunks, We further assign them to different primer pairs based on which file they belong to. Chunks from the same file are assigned to the same primer pairs so that all data chunks of a file in a tube will be sequenced out in one shot. Consequently, retrieving a file needs sequencing with at most the number of total DNA tubes. The remaining chunks in the other tubes will start another round of initial clustering and refinement until all data chunks are allocated.

The time complexity of initial clustering consists of two parts: 1) calculating merge priority between any two chunks and 2) recalculating merge priority after each merging. The time complexity of the first calculation of merge priority between every two chunks is O($n^2$). There are about $\log_n$ to n times of merging, each merging needs an O(n) recalculation. Therefore, the overall time complexity of the initial clustering is O($n^2$). The time complexity of refinement is O(n) as it calculates the merge priority between all other chunks and the selected cluster. Since there are constant times of iteration of initial clustering and refinement (i.e., the number of tubes required), the overall time complexity of the collision aware data allocation scheme is O($n^2$).

\section{Evaluation}
\label{sec:multitube evaluation}
This section evaluates the capacity improvements of the collision aware data allocation scheme. Following previous works~\cite{organick2018random}\cite{wei2022dna}, we generate a primer library with 28,000 primers for the evaluation. To obtain DNA tube capacity, we adopt the 1.55 $\times 10^6$ as the parallel factor and apply the same capacity calculation as described in Section~\ref{sec:tube capacity}. We also test the trade-off for different data chunk sizes including average number of sequencing to retrieve a file and execution time of the allocation.

\subsection{Capacity Improvements with Different Encoding Schemes}
\label{sec:capacity improvement}
In this subsection, we use three state-of-the-art encoding schemes to encode digital data and apply the collision aware data allocation scheme to evaluate the improvements in the number of usable primers and overall storage capacity. 

To fairly evaluate the collision aware data allocation scheme, we designed two baselines for comparison:
\begin{itemize}	
    \item \textbf{Sequential allocation:} sequentially assign data chunks into tubes one after another.
    	
    \item \textbf{UPGMA clustering:} The same procedure as our collision aware data allocation except the merge priority is replaced with UPGMA (Section~\ref{sec:initial clustering}). 

    \item \textbf{Collision aware data allocation:} Our proposed scheme, including initial clustering and refinement.
\end{itemize}	

We have collected digital data from \textit{ImageNet}~\cite{imagenet_cvpr09}, \textit{LibriSpeech}~\cite{panayotov2015librispeech} and \textit{InternetArchive}~\cite{InternetArchive}. The types of digital data have insignificant influence on collisions because all possible bit sequences will appear multiple times as data set scales~\cite{wei2022dna}. Therefore, we mixed the above data and used AES encryption on the input data to generate "random" input which mimics a common archival storage input. When encoding, we first apply Reed–Solomon(255,239) on the input digital data to append 16 parity bytes to each 239 data bytes. We then implemented four state-of-the-art encoding schemes to encode the data respectively. The encoding schemes we used are:
\begin{itemize}	
    \item \textbf{CAC}~\cite{wei2024} is a collision aware encoding scheme. Each time, it encodes a binary triplet to a DNA triplet based on a predefined encoding table. Every possible binary triplet has 4 DNA triplets as candidates. CAC selects a proper candidate based on the previous 17 encoded DNA bases. It tries to select the candidate that introduces no homopolymer, less complementary sequence, and more balanced GC content. This enables the payloads dissimilar to many primers and thus avoid collisions with many primers.
   
    \item \textbf{Rotation} code ~\cite{bornholt2016dna}\cite{organick2018random}\cite{li2021img}\cite{goldman2013towards} converts binary data to ternary. Then it encodes each ternary digit to a DNA base depending on the last encoded base. The current ternary digit must be encoded to a base different from the last base. The encoding density is 1.58 bits/base. 
    
    \item \textbf{Blawat}~\cite{blawat2016forward} encodes every eight binary bits to five bases. Blawat code selectively encodes bases 3 \& 5 to ensure bases 1 \& 2 \& 3 will not form a homopolymer, and bases 5 \& 6 will not be identical. The encoding density is 1.6 bits/base. 

    \item \textbf{Grass}~\cite{grass2015robust} first converts binary bits to Galois Field of size 47 (i.e., GF(47)). Every GF(47) digit is mapped to a base triplet in which the second base differs from the third base. The longest homopolymer length is three. The encoding density is 1.77 bits/base.
\end{itemize}

\begin{figure}[htbp]
    \centering
        \includegraphics[width=0.85\textwidth,height=5.7cm]{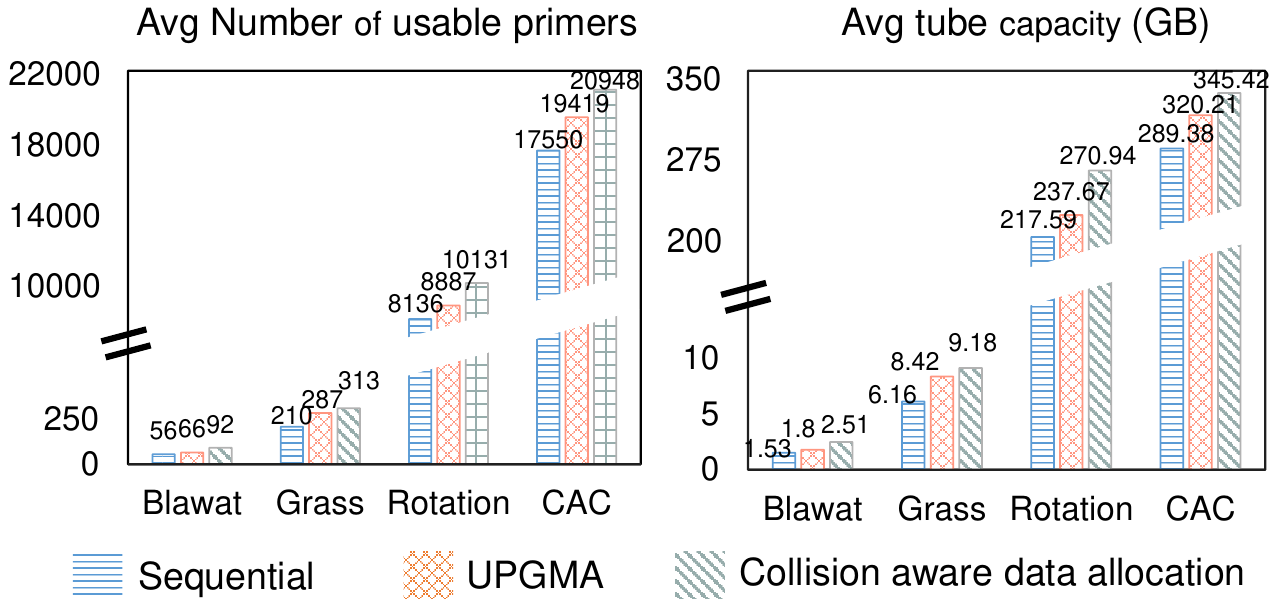}
        \caption{Enhancements of collision aware data allocation in the average number of usable primers per tube and average storage capacity per tube (total five DNA tubes).}
        \label{fig:multitube_evaluation1}
\end{figure}

We respectively performed the two baselines and the collision aware data allocation scheme on the DNA sequences encoded by the four encoding schemes until we fully filled five DNA tubes~\footnote{The two clustering scheme actually determines the number of tubes according to the size of input data, and the last tube may have spare space. For a fair comparison, we conduct multiple experiments and find the input data size that just fills the five tubes.}. The results are shown in Figure~\ref{fig:multitube_evaluation1}. Compared with sequential data allocation, collision aware data allocation can increase thousands of the average number of usable primers per tube for Rotation code (e.g., from 8136 to 10131) and CAC (from 17550 to 20948). This increase in the number of usable primers can enhance the average tube capacity from 217 GB to 270.94 GB and from 289 GB to 345 GB, roughly a 25\% improvement and a 20\% improvement. However, the increase in the number of usable primers and tube capacity is insignificant in Blawat Code and Grass Code (i.e., an increase in tens of usable primers and several GB of tube capacity). This is mainly because the four encoding schemes originally had very different numbers of collisions and collided primers. CAC and Rotation code initially have more usable primers than Blawat and Grass (i.e., CAC: 17550, Rotation: 8,136, Blawat: 56, Grass: 210 in sequential data allocation). That is because CAC and Rotation code create a particular sequence pattern on their payloads that is dissimilar to primers. For example, every DNA base in payloads of Rotation code will differ from the last base. This pattern makes the Rotation code avoid collisions with primers that have several identical consecutive bases. In comparison, Blawat code and Grass code do not restrict a string pattern on their payloads and thus suffer more collisions (see more analysis in ~\cite{wei2022dna}). This difference makes DNA payloads of Blawat code and Grass code have more collisions and more collided primers. In the aspect of data allocation, each data chunk will have much more collided primers if it is encoded by Blawat and Grass. In this case, it is harder to partition data chunks into several groups with independent sets of collided primers. According to our evaluation, the collided primers that existed in all five tubes occupied more than 95\% total primers in the primer library even after the collision aware data allocation. As a consequence, most allocation schemes including the collision aware data allocation scheme can barely increase the number of usable primers and capacity for Blawat code and Grass code.

When comparing the collision aware data allocation scheme with UPGMA based Clustering, the collision aware data allocation scheme can improve more in terms of the number of usable primers and tube capacity. As discussed in Section~\ref{sec:initial clustering}, during clustering, UPGMA calculates an average distance based on the distance of all pairs of data chunks in two clusters. Although comprehensive, this distance calculation does not fit our goal: find clusters that each cluster has a small union of collided primers.  The collision aware data allocation uses the union and intersection of different cluster's collided primers as the distance criteria (i.e., merge priority, Section~\ref{sec:initial clustering}). We always merge clusters with a smaller union and a bigger intersection in terms of collided primers. Since we adopted hierarchical clustering, the merge process keeps monitoring the merge priority among clusters instead of directly clustering upon data chunks. As a result, the collision aware data allocation performs better than UPGMA.

The execution time of the data allocation scheme is acceptable compared with DNA synthesis and sequencing. The scheme takes 56 minutes to allocate 30GB of data on a machine with 32 AMD EPYC 7302P processors, 128GB DRAM, and 20TB SSD. There is no public record of large-scale DNA synthesis and sequencing throughput for DNA storage. According to the information of the biology community, the most advanced DNA sequencing (Illumina Sequencing) takes hours to one day to process 30GB of data. And DNA synthesis requires at least days to process the data.

\subsection{Trade-offs of Different chunk sizes}
In this section, we investigate the trade-offs with different chunk sizes. Chunk size is the granularity of data allocation, different sizes of chunks will have different numbers of collided primers per chunk. This brings to different capacity improvement and also different numbers of sequencing to retrieve a file as the file may be splitted into different numbers of chunks and distributed to different tubes. We adopt the analysis based on CAC as it is the most collision resistant encoding scheme. We adjust chunk size to 1KB, 4KB, 16KB, 256KB, and 1MB for the investigation.

\begin{figure}[htbp]
    \centering
        \includegraphics[width=0.95\textwidth,height=4.8cm]{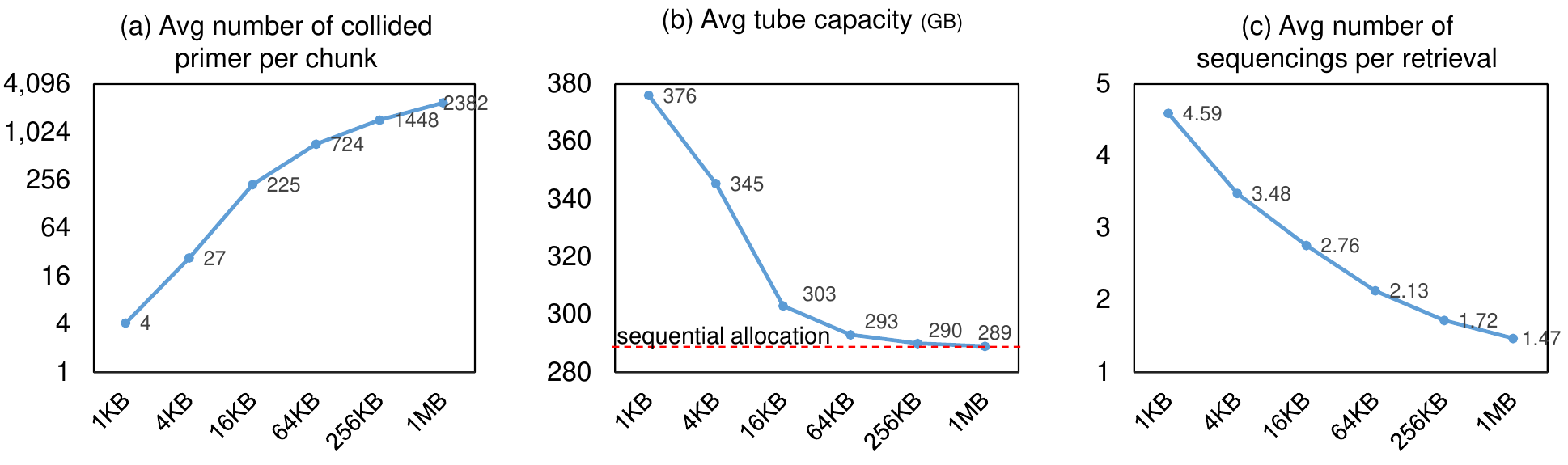}
        \caption{The trade-offs with different chunk sizes: (a) The average number of collided primers per chunk; (b) The average tube capacity if apply the collision aware data allocation; (c) The average number of sequencings required to retrieve a file when data chunk size changes}
        \label{fig:multitube_evaluation2}
\end{figure}

Figure~\ref{fig:multitube_evaluation2} shows the trade-offs with different chunk sizes. First, as chunk size increases, the average number of collided primers per chunk increases. The average number of collided primers per chunk increases faster when chunk size is small (e.g., from 1KB to 4KB and 16KB). When chunk size increases from mid to large (e.g., from 256KB to 1MB), the increase of collided primers slows. This may be because the common collide primers that appear frequently have already been counted when chunk size is small and the remaining collided primers may appear infrequently. The increase of collided primers per chunk significantly affects the data allocation. The average tube capacity (of five tubes) decreases greatly when the chunk size increases. The coarse allocation granularity causes difficulty in allocation as each big data chunk has many more collide primers than small data chunks. Specifically, when the chunk size increases to 256KB and 1MB, the average tube capacity is almost the same as the sequential allocation. Only chunk sizes of 4KB and 1KB can see a desirable improvement in tube capacity.

On the other hand, bigger chunk size has benefits on data retrieval as a file is cut into a fewer number of chunks. The collision aware data allocation scheme can distribute one file's chunks into multiple different tubes. In this case, the retrieval of this file needs multiple DNA sequencing respectively performed on multiple DNA tubes. We record the data chunk distribution of 1000 random files and calculate the average number of sequencing required to retrieve a file. For data chunks allocated in one tube, we assume they will be put in the same primer pair so that only one DNA sequencing is required unless the file's chunks are too large to put into one primer pair (the capacity of the primer pair is determined by encoding density. For CAC, one primer pair can accommodate 36MB of data). The result is shown in Figure~\ref{fig:multitube_evaluation2}(c). As the chunk size increases, the average number of sequencing needed to retrieve a file decreases. It decreases slower when chunk size is large (e.g., from 256KB to 1MB). That is mainly because a considerable portion of files (in our case, more than 40\%) are smaller than 256KB (e.g., documents). The small files are also common in a typical archival system. The average number of sequencing per retrieval is capped below 5 even for 1KB chunk size because our evaluation is based on five DNA tubes.

Considering the improvement in tube capacity and the overhead on the number of sequencing per retrieval, a small chunk size like 1KB or 4KB will be desirable. However, considering the exponential growth in terms of clustering execution time (e.g., for 1KB chunk size, 30GB input data takes more than 16 hours to allocate while 4KB needs fewer than 1 hour), we finally select 4KB as the chunk size. Besides, the overhead of a small chunk size also includes extra index space to record the mapping between files and chunks. This overhead is highly data-related (e.g., portions of small and big files) and is out of this paper's scope.

\section{Conclusion}\label{sec:multitube conclusion}
In this paper, we find the primer-payload collisions will not only affect single tube capacity but also multi-tube capacity. Even if a collision-resistant encoding scheme like CAC is used, the remaining collision will still spread over almost all tubes and cause repeated capacity reduction. We propose a pre-processing method called collision aware data allocation to remedy the influence of collisions. The allocation scheme limited the spread of collision so that primers disabled in one tube are still reusable in other tubes. The evaluation shows the data allocation scheme can effectively increase overall storage capacity. We further discuss the trade-offs of different allocation granularity and select 4KB as the proper chunk size. In the future, we would further investigate the potential enhancement of the allocation speed to reduce the execution time required for the data allocation.

\begin{acks}

\end{acks}

\bibliographystyle{ACM-Reference-Format}
\bibliography{ref}

\end{document}